\documentclass{PoS}

\title{Production of charged Higgs boson pairs \\in the $pp \to ppH^{+}H^{-}$ reaction at the LHC and FCC}

\ShortTitle{$pp \to ppH^{+}H^{-}$ reaction at the LHC and FCC}

\author{\speaker{Piotr Lebiedowicz}\thanks{This research was partially supported 
by the START fellowship from the Foundation for Polish Science, 
the MNiSW Grant No. IP2014~025173 (Iuventus Plus),
and by the Polish National Science Centre Grant No. 2014/15/B/ST2/02528 (OPUS).}
and {Antoni Szczurek\thanks{Also at University of Rzesz\'ow, 35-959 Rzesz\'ow, Poland.}}\\
        The H. Niewodnicza\'{n}ski Institute of Nuclear Physics Polish Academy of Sciences,\\
        ul. Radzikowskiego 152, 31-342 Krak\'ow, Poland\\
        E-mail: \email{Piotr.Lebiedowicz@ifj.edu.pl},\\
        E-mail: \email{Antoni.Szczurek@ifj.edu.pl}}

\abstract{We present differential cross sections for the $pp \to ppH^{+}H^{-}$ reaction
via photon-photon fusion with exact kinematics. 
We show predictions for $\sqrt{s}$ = 14 TeV (LHC) 
and at the Future Circular Collider (FCC) for $\sqrt{s}$ = 100 TeV. 
The integrated cross section for $\sqrt{s}$ = 14~TeV (LHC) is about 0.1~fb 
and about 0.9~fb at the FCC for $\sqrt{s}$ = 100~TeV when assuming $m_{H^{\pm}} = 150$~GeV.
We present distributions in diHiggs boson invariant mass. 
The results are compared with those obtained within equivalent-photon approximation. 
We discuss also first calculations of cross section 
for exclusive diffractive pQCD mechanism with estimated limits on the $g_{hH^{+}H^{-}}$
coupling constant within 2HDM based on the LHC experimental data. 
The diffractive contribution is much smaller than the $\gamma \gamma$ one. 
Absorption corrections are calculated differentially for various distributions. 
In general, they lead to a damping of the cross section. 
The damping depends on $M_{H^{+}H^{-}}$ invariant mass 
and on four-momentum transfers squared in the proton line. 
We discuss a possibility to measure the exclusive production 
of $H^{\pm}$ bosons.}

\FullConference{The European Physical Society Conference on High Energy Physics\\
		22--29 July 2015\\
		Vienna, Austria}

\begin{document}

\section{Introduction}

There are extensive phenomenological studies 
of exclusive processes in search for effects beyond the Standard Model.
The Higgs sector in both the MSSM and 2HDM contains five states:
three neutral [two $CP$-even ($h$, $H$) and one $CP$-odd ($A$)]
and two charged ($H^{+}$, $H^{-}$) Higgs bosons.
In general, either $h$ or $H$ could correspond to the SM Higgs.
Discovery of the heavy Higgs bosons of the Minimal Supersymmetric Standard Model (MSSM) 
or more generic Two-Higgs Doublet Models (2HDMs)
poses a special challenge for future colliders.
One of the international projects currently under consideration
is the Future Circular Collider (FCC) \cite{FCC}.

The main advantage of exclusive processes is that background contributions 
are strongly reduced compared to inclusive processes.
A good example are searches for exclusive production of supersymmetric Higgs boson
\cite{Tasevsky:2014cpa}, anomalous boson couplings 
for $\gamma \gamma \to W^+ W^-$ \cite{anom_WW}
or for $\gamma \gamma \to \gamma \gamma$ \cite{anom_AA}.
So far these processes are usually studied in the so-called 
equivalent-photon approximation (EPA), see e.g.~\cite{Budnev:1974de}.
Within the Standard Model the cross section for the $p p \to p p W^+ W^-$
reaction is about 100~fb at $\sqrt{s} = 14$~TeV \cite{Lebiedowicz:2012gg}. 
The exclusive two-photon induced reactions could be also used in searches for neutral 
technipion in the diphoton final state \cite{Lebiedowicz:2013fta}.
Gluon-induced processes could also contribute to the exclusive
production of $W^+ W^-$ \cite{Lebiedowicz:2012gg}, $W^{\pm} H^{\mp}$ \cite{Enberg:2011qh},
$H^{+} H^{-}$ \cite{Lebiedowicz:2015cea} via quark loops.
However, the corresponding cross sections are rather small
mainly due to suppression by Sudakov form factors and the gap survival factor.

\section{Formalism}

\begin{figure}
\begin{center}
\includegraphics[width=0.25\textwidth]{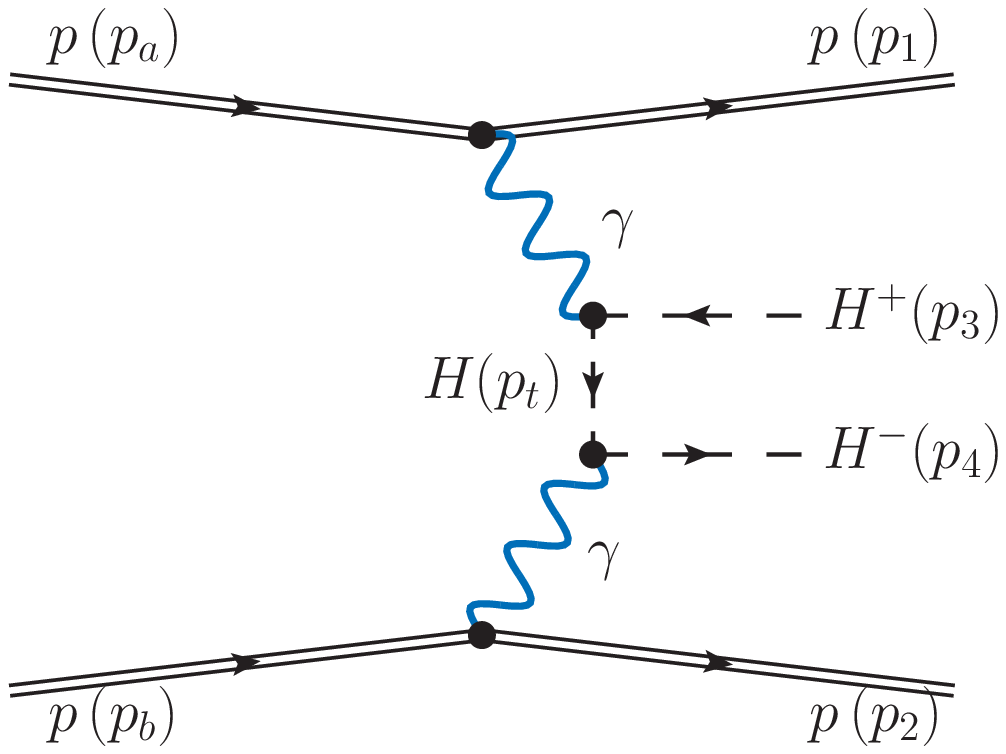}
\includegraphics[width=0.25\textwidth]{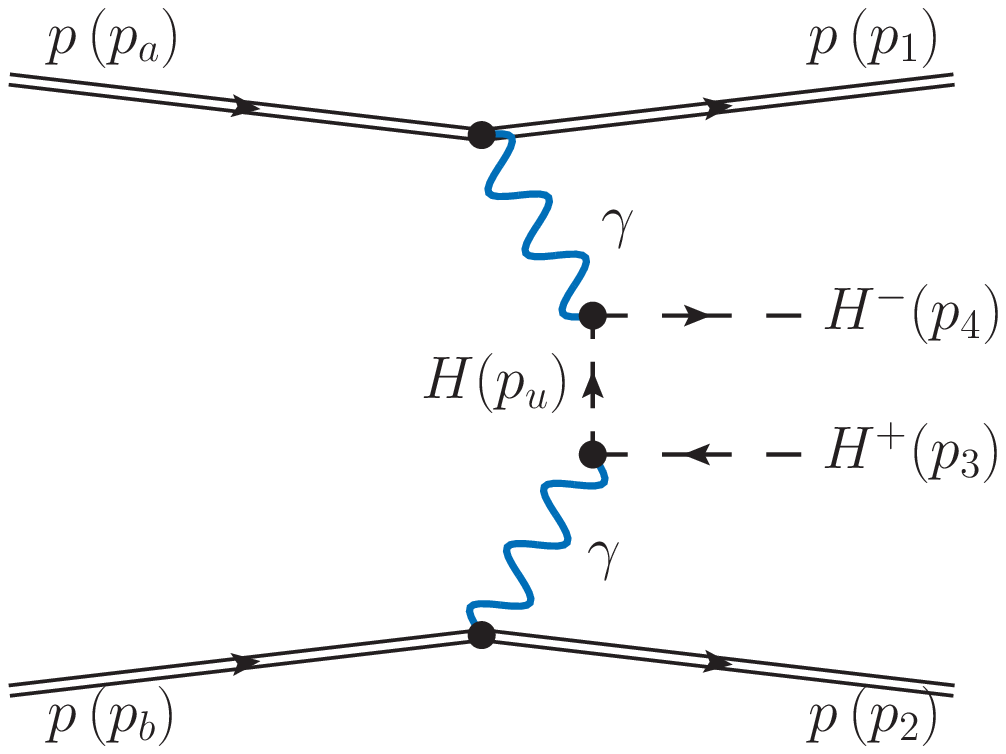}
\includegraphics[width=0.25\textwidth]{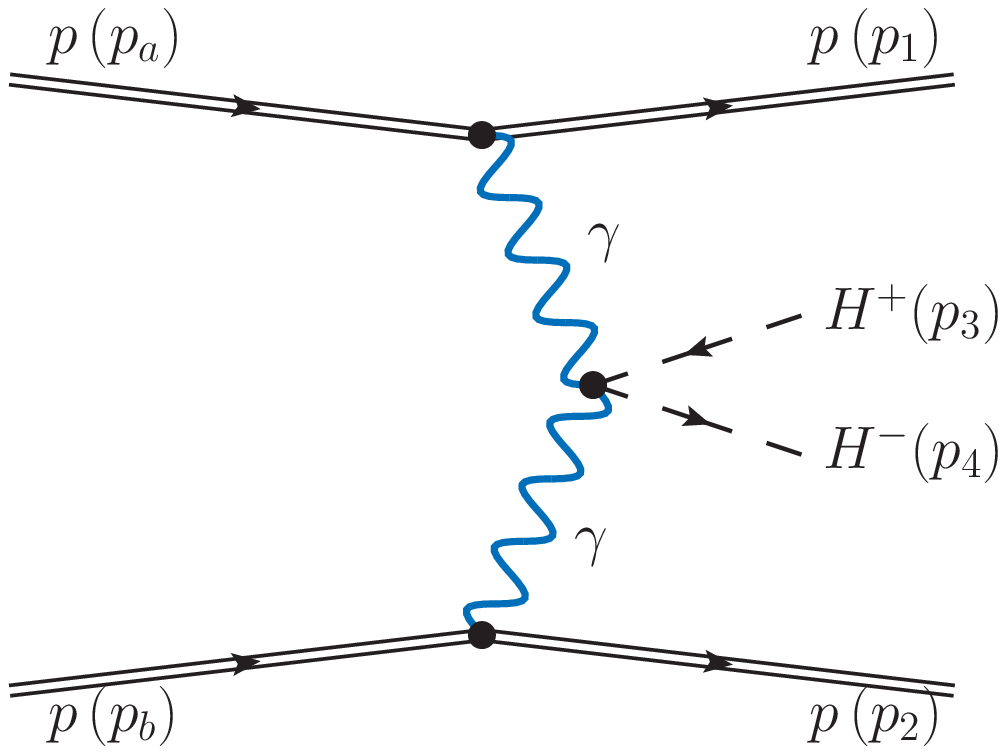}
\caption{Born diagrams for exclusive production of $H^{+}H^{-}$ pairs in $pp$-collisions via $\gamma \gamma$ exchanges.}
\label{fig:1}
\end{center}
\end{figure}
We consider the exclusive production of charged Higgs bosons discussed recently in \cite{Lebiedowicz:2015cea}
\begin{eqnarray}
p(p_{a},\lambda_{a}) + p(p_{b},\lambda_{b}) \to
p(p_{1},\lambda_{1}) + H^{+}(p_{3}) + H^{-}(p_{4}) + p(p_{2},\lambda_{2}) \,,
\label{2to4_reaction}
\end{eqnarray}
where $p_{a,b}$, $p_{1,2}$ and $\lambda_{a,b}$, 
$\lambda_{1,2} = \pm \frac{1}{2}$ 
denote the four-momenta and helicities of the protons, 
and $p_{3,4}$ denote the four-momenta of the charged Higgs bosons, 
respectively.
In general, the cross section 
for the considered exclusive $2 \to 4$ process (\ref{2to4_reaction})
can be written as
\begin{eqnarray}
d \sigma = 
\frac{(2 \pi)^{4}}{2s} 
{\overline{|{\cal M}_{pp \to pp H^+ H^-}|^2}}
\frac{d^3 p_1}{(2 \pi)^{3} 2 E_1} 
\frac{d^3 p_2}{(2 \pi)^{3} 2 E_2}
\frac{d^3 p_3}{(2 \pi)^{3} 2 E_3} 
\frac{d^3 p_4}{(2 \pi)^{3} 2 E_4} 
\delta^{4} \left(E_a + E_b -p_1 - p_2 - p_3 - p_4 \right)\,, \qquad
\label{differential_cs}
\end{eqnarray}
where energy and momentum conservations have been made explicit
\footnote{The details on how to conveniently reduce the number of kinematic 
integration variables are discussed in \cite{Lebiedowicz:2009pj}.
Above $\overline{|{\cal M}|^2}$ is the $2 \to 4$ amplitude squared 
averaged over initial and summed over final proton polarization states.
The phase space integration variables are taken the same as 
in \cite{Lebiedowicz:2009pj}, except that proton transverse momenta 
$p_{1t}$ and $p_{2t}$ are replaced by
$\xi_1$ = log$_{10}(p_{1t}/p_{0t})$ and 
$\xi_2$ = log$_{10}(p_{2t}/p_{0t})$, respectively, where $p_{0t}$ = 1 GeV.}.

The full amplitude for the $pp \to pp H^{+} H^{-}$ reaction is
\begin{eqnarray}
{\cal {M}}_{pp \to pp H^{+} H^{-}} =
{\cal {M}}_{pp \to pp H^{+} H^{-}}^{\mathrm{Born}} + 
{\cal {M}}_{pp \to pp H^{+} H^{-}}^{\mathrm{absorption}}\,,
\label{amp_full}
\end{eqnarray}
where the Born amplitudes via $\gamma \gamma$ exchanges (see diagrams of Fig.~\ref{fig:1}) are calculated as
\begin{eqnarray}
{\cal M}^{Born}_{\lambda_a \lambda_b \to \lambda_1 \lambda_2 H^+ H^-}(t_{1},t_{2}) =
V_{\lambda_a \to \lambda_1}^{\mu_1}(t_{1})
D_{\mu_1 \nu_1}(t_1) 
V_{\gamma \gamma \to H^+ H^-}^{\nu_1 \nu_2}
D_{\nu_2 \mu_2}(t_2)
V_{\lambda_b \to \lambda_2}^{\mu_2}(t_{2}) \,, \quad
\label{born}
\end{eqnarray}
where $D_{\mu \nu}(t) = -i g_{\mu \nu}/t$ is the photon propagator.
The $\gamma pp$ vertex takes the form
%
\begin{eqnarray}
V_{\lambda \to \lambda'}^{(\gamma pp) \mu}(t) = e \, \bar{u}(p',\lambda') 
\left( 
\gamma^{\mu} F_{1}(t) + \frac{i}{2 m_{p}} \sigma^{\mu \nu} (p'-p)_{\nu} F_{2}(t)
\right) u(p,\lambda) \,,
\label{vertex_spinors}
\end{eqnarray}
where $u(p,\lambda)$ is a Dirac spinor and $p, \lambda$ and $p', \lambda'$ are initial and final four-momenta and helicities of the protons, respectively.
In the high-energy approximation
one gets the simple formula
%
\begin{eqnarray}
V_{\lambda \to \lambda'}^{(\gamma pp) \mu}(t) \simeq e
\left( \frac{\sqrt{-t}}{2 m_p} \right)^{| \lambda' - \lambda |} F_{i}(t) (p'+p)^{\mu} \,.
\label{vertex}
\end{eqnarray}
%
The tensorial vertex in Eq.~(\ref{born}) for the 
$\gamma \gamma \to H^+H^-$ subprocess
is a sum of three-level amplitudes corresponding to 
$t$, $u$ and contact diagrams of Fig.~\ref{fig:1},
\begin{eqnarray}
V_{\gamma \gamma \to H^+ H^-}^{\nu_1 \nu_2} =
i e^{2} \frac{(q_{2} - p_{4} + p_{3})^{\nu_1} (q_{2} - 2p_{4})^{\nu_2}}{p_{t}^{2} - m_{H}^{2}}
+ i e^{2} \frac{(q_{1} - 2p_{4})^{\nu_1} (q_{1} - p_{4} + p_{3})^{\nu_2}}{p_{u}^{2} - m_{H}^{2}}  
- 2 i e^{2} g^{\nu_1 \nu_2} \,, \qquad
\label{central_vertex}
\end{eqnarray}
where $p_{t}^{2} = (q_{2} - p_{4})^{2} = (q_{1} - p_{3})^{2}$
and $p_{u}^{2} = (q_{1} - p_{4})^{2} = (q_{2} - p_{3})^{2}$.

The amplitude including $pp$-rescattering corrections 
between the initial- and final-state protons
in the four-body reaction discussed here can be written as
\begin{eqnarray}
{\cal M}_{\lambda_a \lambda_b \to \lambda_1 \lambda_2 H^+ H^-}^{\mathrm{absorption}}(s,\textbf{p}_{1t},\textbf{p}_{2t})=&&
\frac{i}{8 \pi^{2} s} \int d^{2} \textbf{k}_{t} \;
{\cal M}_{\lambda_{a}\lambda_{b} \to \lambda'_{a}\lambda'_{b}}(s,-\textbf{k}_{t}^{2})\;
{\cal M}_{\lambda'_{a}\lambda'_{b}\to \lambda_{1}\lambda_{2} H^{+} H^{-}}^{\mathrm{Born}}(s,\tilde{\textbf{p}}_{1t},\tilde{\textbf{p}}_{2t})\,, \qquad
\label{abs_correction}
\end{eqnarray}
where $\tilde{\textbf{p}}_{1t} = {\textbf{p}}_{1t} - {\textbf{k}}_{t}$ and
$\tilde{\textbf{p}}_{2t} = {\textbf{p}}_{2t} + {\textbf{k}}_{t}$.
Here ${\textbf{p}}_{1t}$ and ${\textbf{p}}_{2t}$
are the transverse components of the momenta of the final-state protons
and ${\textbf{k}}_{t}$ is the transverse momentum carried by additional pomeron exchange.
${\cal M}_{pp \to pp}(s,-{\textbf{k}}_{t}^{2})$ 
is the elastic $pp$-scattering amplitude
for large $s$ and with the momentum transfer $t=-{\textbf{k}}_{t}^{2}$.

The photon induced processes are treated usually in the equivalent-photon
approximation (EPA) in the momentum space, see e.g.
\cite{Lebiedowicz:2012gg,Lebiedowicz:2013fta,Lebiedowicz:2015cea}.
\footnote{An impact parameter EPA was considered recently in \cite{Dyndal:2014yea}.
Only very few differential distributions can be obtained in the EPA approach.}
In this approximation, when neglecting photon transverse momenta, 
one can write the differential cross section as
\footnote{An approach including transverse momenta of photons was discussed 
recently in \cite{daSilveira:2014jla}.}
\begin{equation}
\frac{d \sigma}{d y_3 d y_4 d^2p_{tH}} = \frac{1}{16 \pi^2 {\hat s}^2} 
x_1  f(x_1) x_2 f(x_2) \overline{|{\cal M}_{\gamma \gamma \to H^+ H^-}|^2} \,,
\label{EPA_formula}
\end{equation}
where ${\hat s} = s x_{1} x_{2}$ and $f(x)$'s are elastic fluxes of the equivalent photons 
(see e.g.\cite{Budnev:1974de}) 
as a function of longitudinal momentum fraction with respect to the parent proton 
defined by the kinematical variables of the charged Higgs bosons:
%
$x_1 = \frac{m_{tH}}{\sqrt{s}}(e^{y_3} + e^{y_4})$,
$x_2 = \frac{m_{tH}}{\sqrt{s}}(e^{-y_3} + e^{-y_4})$,
$m_{tH} = \sqrt{|\vec{p}_{tH}|^{2}+m_{H}^{2}}$.
%
$\overline{|{\cal M}|^2}$ is the $\gamma \gamma \to  H^+ H^-$ amplitude squared 
averaged over the $\gamma$ polarization states.

\section{Results}

\begin{figure} 
\center
\includegraphics[width=0.49\textwidth]{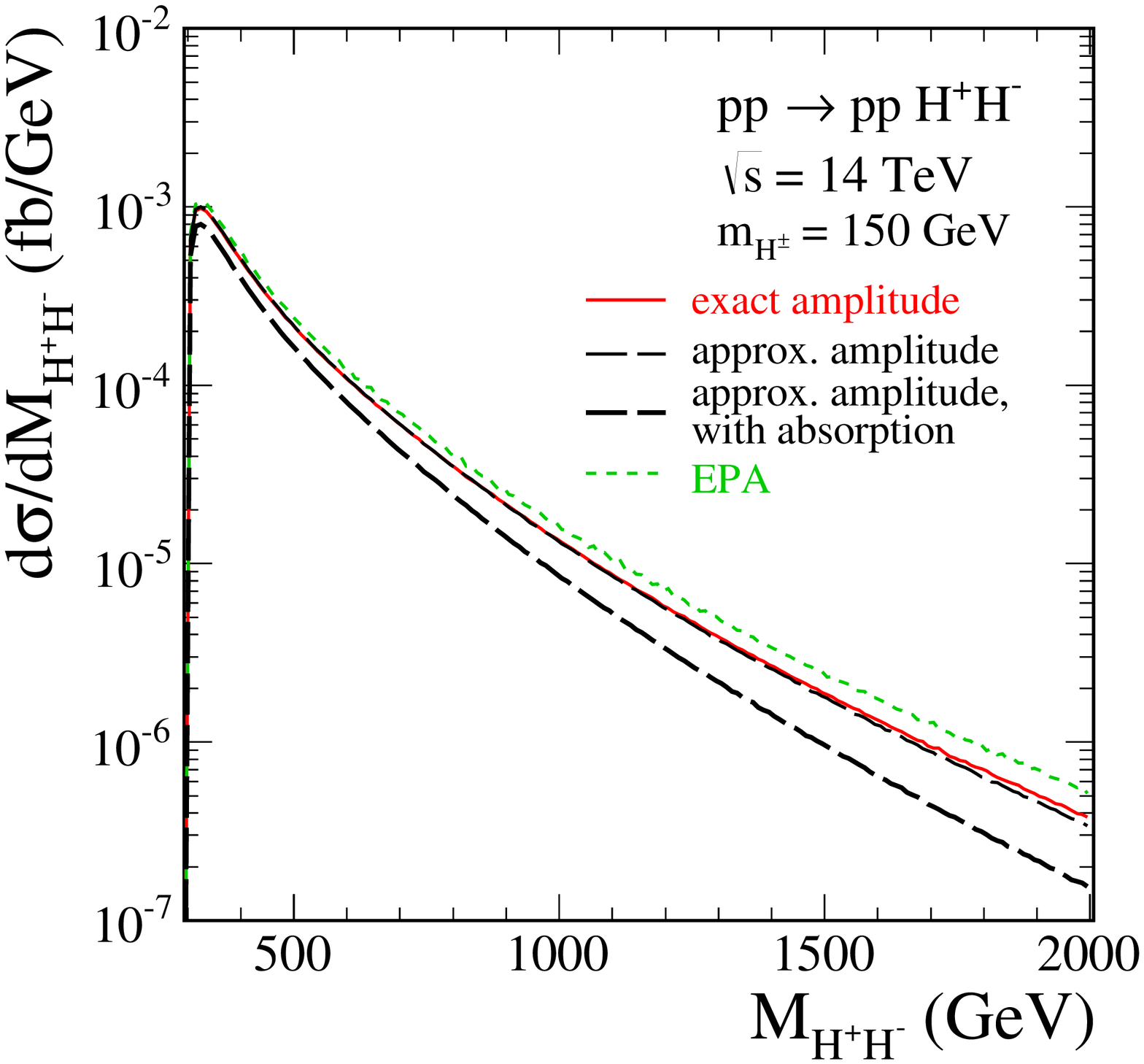}
\includegraphics[width=0.49\textwidth]{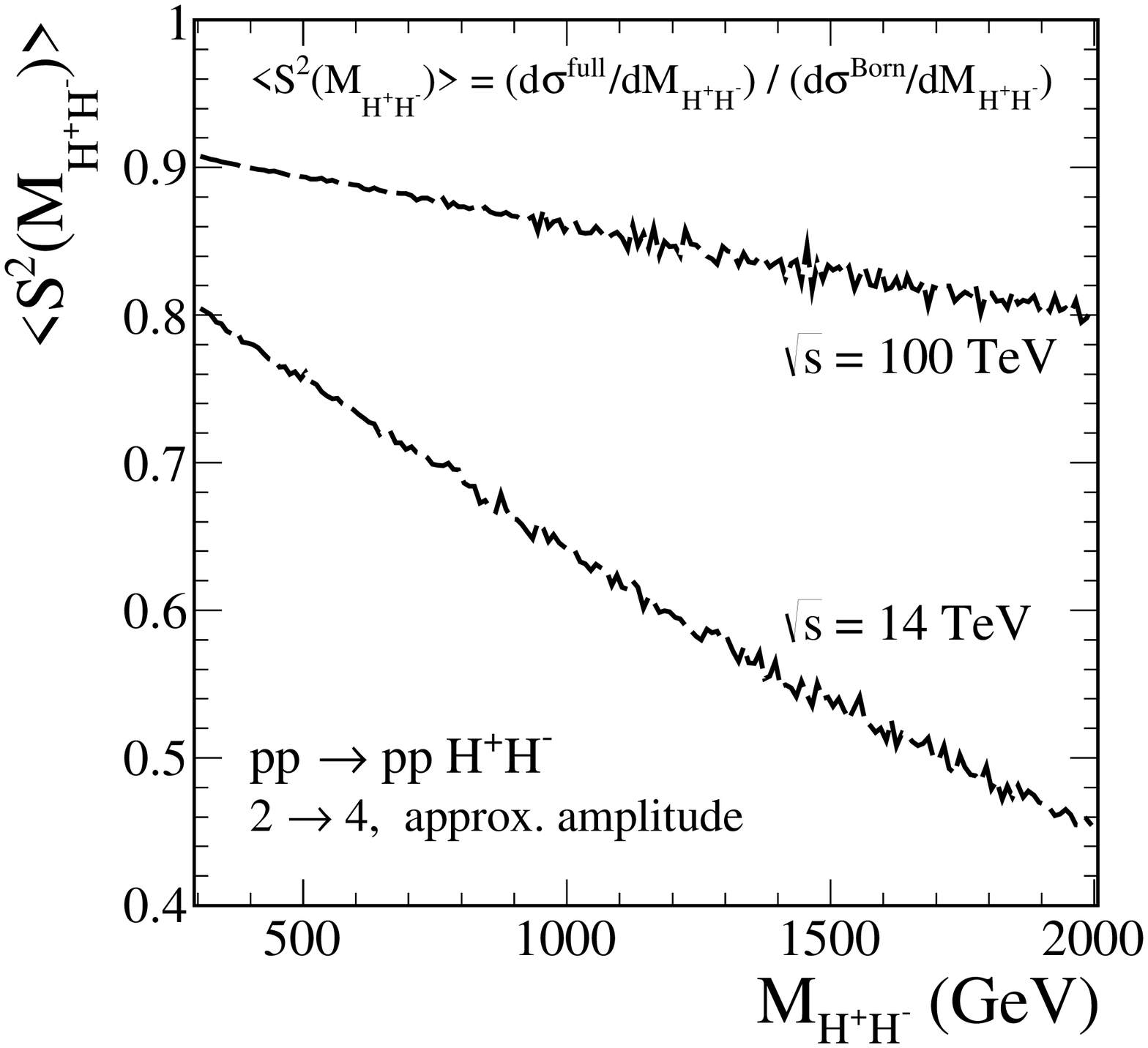}
\caption{
In the left panel we show diHiggs boson invariant mass distribution at $\sqrt{s} = 14$~TeV.
The red solid line represents the calculation for exact $2 \to 4$ kinematics 
and amplitude (including spinors of protons, see Eq.~(2.5)).
The black upper and lower long-dashed lines correspond to calculations 
in the high-energy approximation Eq.~(2.6)
without and with the absorption corrections, respectively.
The green short-dashed line represents results of EPA, see Eq.~(2.9).
In the right panel we show the dependence of the gap survival factor due to $pp$ interactions 
on $M_{H^{+}H^{-}}$ for exact $2 \to 4$ kinematics for the LHC and FCC energies.
}
\label{fig:2}
\end{figure}

In Fig.~\ref{fig:2} we show invariant mass distribution
of the $H^+ H^-$ system in a broad range of the invariant masses.
In the left panel we compare results for the exact kinematics 
and for the EPA calculations.
In contrast to inclusive processes, the exclusive reaction (\ref{2to4_reaction})
is free of the model parameter uncertainties,
at least in the leading order, except of the mass of the Higgs bosons.
In the right panel we show the dependence of absorption on $M_{H^{+}H^{-}}$.
This is quantified by the ratio of full (with the absorption corrections) and Born 
differential cross sections $\langle S^{2}(M_{H^{+}H^{-}})\rangle$.
The values of the gap survival factor $\langle S^{2}\rangle$
for different masses of $H^{\pm}$ bosons $m_{H^{\pm}} = 150, 300, 500$~GeV
are, respectively, $0.77, 0.67, 0.57$ for $\sqrt{s} = 14$~TeV (LHC)
and $0.89, 0.86, 0.82$ for $\sqrt{s} = 100$~TeV (FCC).
In contrast to diffractive processes, the larger the collision energy, 
the smaller the effect of absorption.

\begin{figure}
\center
\includegraphics[width=0.49\textwidth]{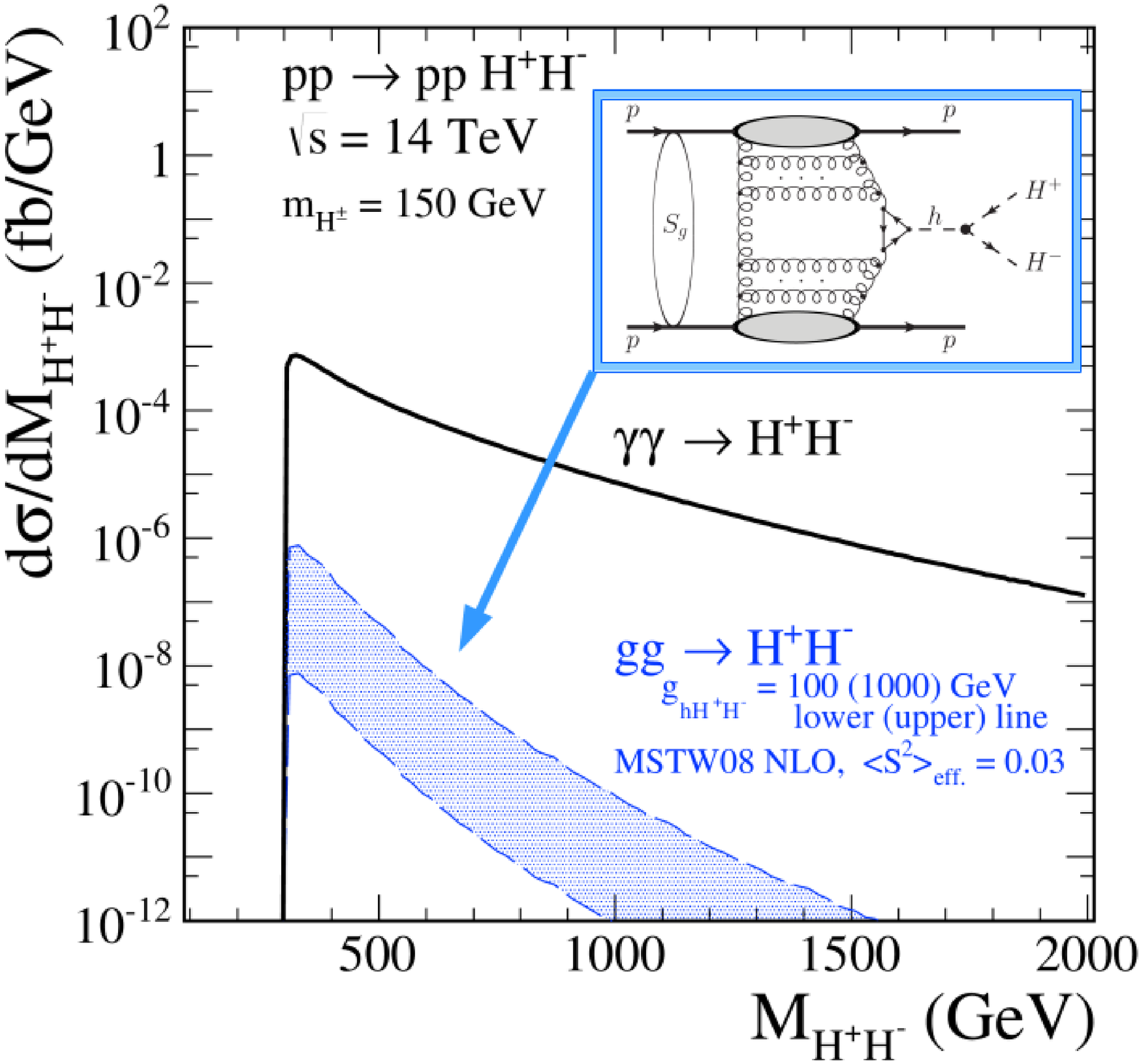}
\includegraphics[width=0.49\textwidth]{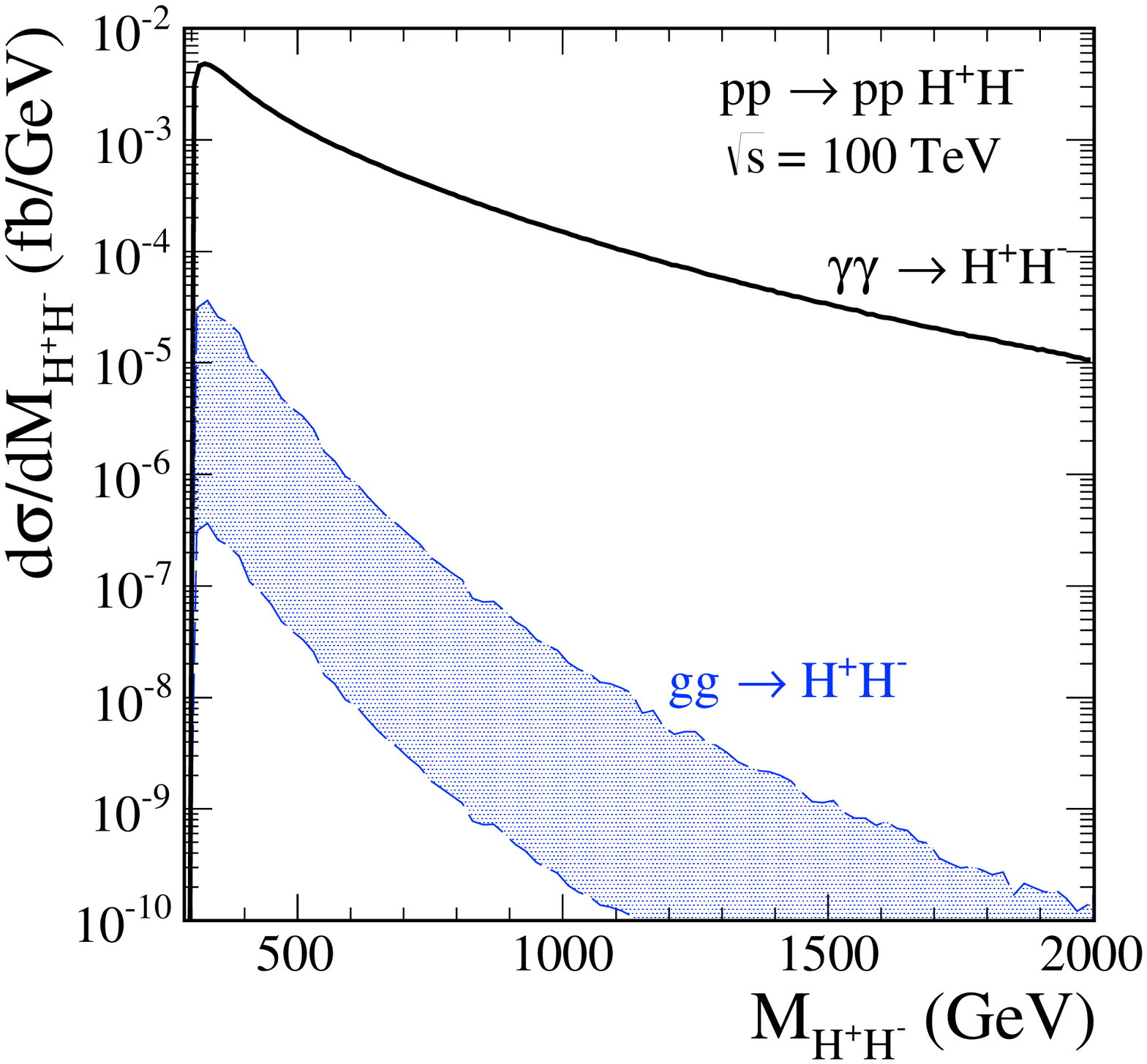}
\caption{
DiHiggs boson invariant mass distributions 
at $\sqrt{s} = 14$~TeV (left panel) and $100$~TeV (right panel).
The short-dashed (online green) lines represent results of EPA.
The upper lines represent the $\gamma \gamma$ contribution.
We also show contribution of the diffractive mechanism (the shaded area)
for the MSTW08 NLO collinear gluon distribution \cite{Martin:2009iq}
and $g_{hH^{+}H^{-}} = 100$ (1000)~GeV for the lower (upper) limit.
}
\label{fig:3}
\end{figure}

So far we have considered a purely electromagnetic process,
the contribution of which is model independent.
In Fig.~\ref{fig:3} we show also corresponding results for the diffractive contribution
for $m_{H^{\pm}} = 150$~GeV
including the ``effective'' gap survival factor $\langle S^{2}\rangle=0.03$
for $\sqrt{s}$ = 14 TeV (left panel) and $\sqrt{s}$ = 100 TeV (right panel).
The $g^{*}g^{*} \to H^{+}H^{-}$ hard subprocess amplitude
for the diffractive KMR mechanism
through the $t$-loop and $s$-channel SM Higgs boson ($h^{0}$) is given by
%
\begin{equation}
V_{g^{*}g^{*} \to H^+ H^-} = 
V_{gg \to h} \, \frac{i}{s_{34} - m_h^2 + i m_{h} \Gamma_{h}} \, g_{h H^+ H^-}
\end{equation}
and enters into ${\cal M}_{pp \to pp H^{+}H^{-}}$
invariant $2 \to 4$ amplitude for the diffractive process as in \cite{Maciula:2010tv, Lebiedowicz:2012gg}.
The triple-Higgs coupling constant $g_{h H^+ H^-}$ is model dependent.
In the MSSM model it depends only on the parameters $\alpha$ and $\beta$. 
In the general 2HDM it depends also
on other parameters such as the Higgs potential $\lambda$-parameters 
or masses of Higgs bosons. 
In addition to the result for the 2HDM set of parameters 
(alignment limit, $\beta - \alpha \approx \pi/2$), 
we also show result with the upper limit $g_{hH^{+}H^{-}} = 1000$~GeV.
The corresponding couplings in the MSSM are smaller than 50~GeV.
One can observe from Fig.~\ref{fig:3} that the cross section for the exclusive diffractive process
is much smaller than that for $\gamma \gamma$ mechanism both for LHC and FCC.

\section{Conclusions}

We have discussed the exclusive $p p \to p p H^+ H^-$ process at the LHC and FCC.
Results of our exact ($2 \to 4$ kinematics) calculations have been 
compared with those for the equivalent-photon approximation for 
observables accessible in EPA. 
We wish to emphasize that some correlation observables
in EPA are not realistic, or even not accessible, 
to mention here only correlations in azimuthal angle 
between the outgoing protons or the charged Higgs bosons.
We have studied the absorption effects due to proton-proton 
nonperturbative interactions.
The absorptive effects lead to a reduction of the cross section.
We have found interesting dependence of the absorption on $M_{H^{+}H^{-}}$. 
The relative effect of absorption is growing with growing $M_{H^{+}H^{-}}$.
We have predicted that the absorption effects for our two-photon-induced
process become weaker at larger collision energy which is in contrast
to the typical situation for diffractive exclusive processes.
Our study shows that an assumption of no absorption or constant
(independent of phase space) absorption, often assumed in the literature
for photon-photon-induced processes, is rather incorrect.

In addition to calculating differential distributions corresponding to 
the $\gamma \gamma$ mechanism we have performed first calculations
of the $H^+ H^-$ invariant mass for the diffractive KMR mechanism. 
We have tried to estimate limits on 
the $g_{h H^+ H^-}$ coupling constant within 2HDM based on recent
analyses related to the Higgs boson discovery.
The diffractive contribution, even with the overestimated
$|g_{h H^+ H^-}|$ coupling constant, gives a much smaller cross section
than the $\gamma \gamma$ mechanism.

Whether the $p p \to p p H^+ H^-$ reaction can be identified at 
the LHC (run 2) or FCC requires further studies including simulations of the $H^{\pm}$ decays. 
Two $H^{\pm}$ decay channels seem to be worth studying
in the case of light $H^{\pm}$:
$H^{\pm} \to \tau^{+} \nu_{\tau} (\tau^{-} \bar{\nu}_{\tau})$ or $H^{\pm} \to c \bar s (\bar c s)$.
The first decay channel may be difficult due to a competition
of the $p p \to p p W^+ W^-$ reaction which can also contribute to the $\tau^+ \tau^-$ channels.
Although the branching fraction $W^{+} \to \tau^+ \nu_{\tau}$ or
$W^{-} \to \tau^-  \bar{\nu}_{\tau}$ is only about $\frac{1}{9}$,
it is expected to be a difficult irreducible background because
of the relatively large cross section for the $p p \to p p W^+ W^-$. 
In the second case (four quark jets), one could measure
invariant masses of all dijet systems to reduce the $W^+ W^-$ background.
In the case of the heavy $H^{\pm}$ Higgs boson,
the $H^{\pm} \to t \bar{b} (\bar{t} b)$ decay can be considered.
In principle, both the $t$ quark and $b$ jet can be measured. 
In contrast to the previous case we do not know about any sizeable
irreducible background.

\end{document}